\documentstyle[psfig,harvard]{mn}

\def\mbstartxt{$M_{\mathrm B} - 5 \log h = -19.5$}
\def\gsim{~\rlap{$>$}{\lower 1.0ex\hbox{$\sim$}}}

\begin{document}

\title{The Dependence of Velocity and Clustering Statistics on Galaxy Properties}
\author[A.~J.~Benson, C.~M.~Baugh, S.~Cole, C.~S.~Frenk and C.~G.~Lacey]{A.~J.~Benson$^{1,3}$, C.~M.~Baugh$^{1,4}$, S.~Cole$^{1,5}$, C.~S.~Frenk$^{1,6}$ and C.~G.~Lacey$^{1,2,7}$ \\
1. Physics Department, University of Durham, Durham DH1 3LE, England. \\
2. SISSA, Astrophysics Sector, via Beirut, 2-4, 34014 Trieste, Italy. \\
3. E-mail: A.J.Benson@durham.ac.uk \\
4. E-mail: C.M.Baugh@durham.ac.uk \\
5. E-mail: Shaun.Cole@durham.ac.uk \\
6. E-mail: C.S.Frenk@durham.ac.uk \\
7. E-mail: lacey@sissa.it \\
}

\maketitle

\begin{abstract}
We use a combination of N-body simulations of the hierarchical
clustering of dark matter and semi-analytic modelling of the physics
of galaxy formation to probe the relationship between the galaxy
distribution and the mass distribution in a flat, cold dark matter
universe with mean density $\Omega_0=0.3$ ($\Lambda$CDM). We find that
the statistical properties of the galaxy distribution in the model, as
quantified by pairwise velocity dispersions and clustering strength,
can be quite different from those displayed by the dark matter. The
pairwise line-of-sight velocity dispersion of galaxies is sensitive to
the number of galaxies present in halos of different mass. In our
model, which is consistent with the observed galaxy number
distribution, the galaxy velocity dispersion is $\sim 40\%$ lower than
that of the dark matter and is in reasonable agreement with the values
measured in the Las Campanas redshift survey by Jing et al. over two
decades in pair separation.  The origin of this offset is statistical
rather than dynamical, and depends upon the relative efficiency of
galaxy formation in dark matter halos of different mass. Although the
model galaxies and the dark matter have markedly different correlation
functions in real space, such biases conspire to cause the redshift
space correlation functions to be remarkably similar to each
other. Thus, although genuinely biased relative to the dark matter on
small scales, the distribution of galaxies as seen in redshift space,
appears unbiased. The predicted redshift-space galaxy correlation
function agrees well with observations. We find no evidence in the
model for a dependence of clustering strength on intrinsic galaxy
luminosity, unless extremely bright galaxies, two magnitudes brighter
than $L_*$, are considered. However, there are significant differences
when model galaxies are selected either by morphology or by colour.
Early type or red galaxies show a much stronger clustering amplitude
than late type or blue galaxies, particularly on small scales, again
in good agreement with observations.
\end{abstract}

\begin{keywords}
large-scale structure of the Universe; galaxies:formation; galaxies:statistics.
\end{keywords} 

\section{Introduction}

Measurements of the clustering and peculiar motions of galaxies can,
in principle, reveal the way in which galaxies are related to the
large-scale distribution of mass in the universe and thus provide
strong tests of models of galaxy formation. In this paper we
investigate three aspects of the galaxy distribution that are
particularly sensitive to the galaxy-mass connection: the distribution
of relative velocities of galaxy pairs, redshift-space distortions in
statistical measures of galaxy clustering, and the dependence of
clustering strength on intrinsic galaxy properties.

Most current discussion of large-scale structure takes place within
the context of the cold dark matter (CDM) theory. It has long been
known, however, that cold dark matter models in which the galaxies are
assumed to have the same statistical distribution as the dark matter
do not account for the relatively low {\it rms} pairwise velocity
dispersion measured in galaxy surveys. Indeed, the first N-body
simulations of the standard CDM model \cite{defw85} gave values 2-3
times higher that those that had been measured for galaxies in the CfA
redshift survey by \citeasnoun{davis83}. This conflict stimulated the
development of the ``high peak'' model of galaxy formation
\cite{bbks86}, in which galaxies are assumed to be more strongly
clustered than, and therefore to be biased tracers of, the dark matter
distribution. In this case, mass fluctuations at the present day and
the associated peculiar velocities are weaker than in an unbiased
model with the same level of galaxy clustering. Davis {\it et al}
showed that galaxies identified with high peaks in the standard
$\Omega_0=1$ CDM cosmology could approximately match the observed
amplitude of galaxy clustering while having a velocity dispersion
distribution only slightly higher than that measured by
\citeasnoun{davis83} (and similar, in fact, to that in an
$\Omega_0=0.2$ CDM model).

The observational determination of the pairwise velocity dispersion
has been problematic.  It is now recognized that this statistic is
very sensitive to the contribution from galaxies in rich clusters,
which were not fairly represented in early redshift surveys
\cite{mjb93,zurek94,mo96,spn97,sdp97}. More recent estimates from
larger surveys, typically sampling volumes on the order of
$10^{6}h^{-3}$Mpc$^{3}$\footnote{We write the Hubble constant as
$H_0=100h$ km s$^{-1}$ Mpc$^{-1}$.}, are more robust and in much
better agreement with one another \cite{marzke95,jing98,arat98}. The
current values are around $\sim 500$~km~s$^{-1}$, almost twice as
large as the original measurement by \citeasnoun{davis83}.

A complementary way to explore the connection between the
distributions of dark matter and galaxies is to look for a dependence
of clustering on galaxy properties such as intrinsic luminosity,
morphology or colour.  There have been many attempts to measure galaxy
clustering as a function of luminosity, often with conflicting
conclusions
\cite{phillipps87,alimi88,davis88,hamilton88,white88,santiago90,iovino93,park94}.
A number of recent measurements of the redshift-space correlation
function of galaxies selected by absolute magnitude suggests a
dependence of clustering strength on luminosity for galaxies brighter
than $L_{*}$ \cite[but see Loveday et al. 1995 and Hoyle et al. 1999
for counter-examples]{benoist96,tadros96,willmer98,guzzo99}.

A dependence of clustering on morphological type is, however, well
established, with elliptical galaxies found to be more strongly
clustered than spiral galaxies as measured both in angular catalogues
\cite{davis76,giovanelli86,iovino93} and in three-dimensional surveys
\cite{mooreqdot94,loveday95}.  This result can be understood as a
reflection of the morphology-density relation, the statement that
elliptical galaxies preferentially inhabit high density regions of the
Universe \cite{dressler80}.  The dependence of clustering on the
colour of galaxies has also been measured in redshift surveys with the
general consensus that red galaxies are more strongly clustered than
blue galaxies \cite{carlberg96,fevre96,willmer98}. This is, of course,
related to the fact that elliptical galaxies are generally redder than
spiral galaxies, and so this effect is due largely to the same
physical processes that drive the morphological dependence of
clustering. This colour dependence is not as clear cut for studies of
the angular correlation function \cite{infante93,landy96}, where the
red and blue populations may have different redshift distributions.

In order to interpret the significance of these observational results,
it is necessary to replace heuristic biasing schemes with physically
motivated models that can realistically incorporate the role of the
environment in the formation and evolution of galaxies.  Semi-analytic
models of galaxy formation \cite{WF91,kgw93,lacey93,coleetal94,sp98}
follow the formation and evolution of galaxies \emph{ab initio},
within the context of cosmological theories of structure formation.
These models have met with significant successes in describing and
predicting observed properties of galaxies, such as faint galaxy
number counts \cite{WF91,kauff94,baugh96a}, properties as a function
of morphological type \cite{kauff96a,kauff96b,baugh96} and the nature
of Lyman break galaxies at high redshift
\cite{baugh98,governato98,kolatt99}.

In their simplest implementation, semi-analytic models can give
limited information about the clustering of galaxies on large scales
\cite{baugh99} by applying the analytic description of the clustering
of dark matter halos developed by \citeasnoun{mowhite96}. A more
powerful technique recently developed consists of incorporating the
semi-analytic prescription into an N-body simulation, whereby the
clustering of galaxies can be determined directly and reliably down to
small scales \cite{kns,kauff98a,kauff98b,meetal,diaferio99}.  We
applied such a technique in an earlier paper \cite{meetal}, and found
that the semi-analytic model of \citeasnoun{coleetal} produced a
galaxy correlation function in a $\Lambda$CDM cosmology (i.e. a CDM
model with $\Omega_0=0.3$ and $\Lambda/3H^{2}_0=0.7$) that matches
remarkably well the real-space correlation function measured from the
APM galaxy survey \cite{cmbapm}.  On small scales, the model galaxies
are less clustered than the dark matter. A similar conclusion was
reached by \citeasnoun{pearce99} for galaxies formed in cosmological
gas dynamics simulations and by \citeasnoun{colin98} for galactic dark
matter halos formed in high-resolution N-body simulations.

In this paper, we combine our semi-analytic model of galaxy formation
with high-resolution N-body simulations to investigate the clustering
properties of galaxies selected according to various criteria.  We
consider clustering in real and redshift-space and show that galaxy
velocity dispersions are significantly lower than those of the dark
matter, in reasonable agreement with the data. Our results for this
particular statistic differ somewhat from those recently obtained by
\citeasnoun{kauff98a} and we explore the reasons for these
differences.  Kauffmann et al. also investigated the dependence of
clustering amplitude on galaxy properties, finding that both early
type and red galaxies were more strongly clustered than the galaxy
population as a whole, in agreement with the work of \citeasnoun{kns}
and with the trends seen in the observational data. However, they
found no evidence in their models for the luminosity dependent
clustering amplitude measured by \citeasnoun{willmer98}. Our results
for these properties are quite consistent with those of
\citeasnoun{kauff98a}, but we compare our models to different
datasets.

The rest of the paper is laid out as follows.  In \S\ref{sec:desc}, we
briefly describe the semi-analytic model of galaxy formation and the
N-body simulation used in this study.  In \S\ref{sec:red} we describe
how we take into account the effects of redshift space distortions and
calculate galaxy velocity dispersions. We examine the model
correlation functions of galaxies selected by luminosity, morphology
and colour and compare these results with the available observational
data in \S\ref{sec:xitypes}.  Finally, in \S\ref{sec:disc} we present
our conclusions.

\section{Description of the numerical technique}
\label{sec:desc}

We implement the semi-analytic model of \citeasnoun{coleetal} in the
``GIF'' N-body simulations, a full description of which may be found
in \citeasnoun{arj98} and \citeasnoun{kauff98a}.  These are
high-resolution dark matter simulations that follow 17 million
particles within a cosmological cubical volume of side $141.3 h^{-1}$
Mpc. In this paper, we focus attention on the $\Lambda$CDM simulation,
which has cosmological parameters, $\Omega _0=0.3$,
$\Lambda/(3H^{2}_{0}) = 0.7$, $h=0.7$, a power spectrum shape
parameter $\Gamma = 0.21$, and normalisation $\sigma _8 = 0.9$ (to
match the local abundance of hot X-ray clusters; White, Efstathiou \&
Frenk 1993, Eke, Cole \& Frenk 1996.)  Dark matter halos are
identified using the friends-of-friends algorithm \cite{defw85} with a
linking length of 0.2 times the mean interparticle separation.  We
consider only halos consisting of ten or more particles, which
\citeasnoun{kauff98a} have shown are stable clusters in the GIF
simulations, and correspond to a lower halo mass limit of $1.4 \times
10^{11} h^{-1} M_{\odot}$.  \citeasnoun{meetal} have shown that galaxy
catalogues produced by the semi-analytic model with this halo mass
limit are complete for galaxies brighter than $M_{B}-5\log h \approx
-18.3$.  The semi-analytic model populates the simulated volume with
approximately 12,000 galaxies brighter than \mbstartxt.

In order to investigate the clustering properties of very bright, low
abundance galaxies we have used a different simulation of a much
larger volume. This was carried out by the VIRGO Consortium and
hereafter we will refer to it as the ``$512^3$'' simulation. It has
identical cosmological parameters to the GIF $\Lambda$CDM simulation,
although the initial conditions were generated using the transfer
function calculated from {\sc cmbfast} \cite{seljak96} rather than the
fitting function of \citeasnoun{be84} which was used for the GIF
simulations. The simulated volume is a cube of side $479 h^{-1}$Mpc,
populated with 134 million particles, giving a particle mass of $6.8
\times 10^{10}h^{-1}M_\odot$. Resolving halos of 10 or more particles
in this simulation results in galaxy catalogues which are complete for
galaxies brighter than $M_{\rm B}-5\log h\approx -20.3$. Thus, we will
only use this simulation for studying the distribution of the very
brightest galaxies.

We have found that the clustering strength of the dark matter in the
$512^3$ simulation is slightly larger than in the GIF simulation at
large pair separations. The ratio of the real-space dark matter
correlation functions in the two simulations, $\xi_{512^3}/\xi_{\rm
GIF}$, is 1.4 at pair separations of $10 h^{-1}$Mpc, decreasing to 1.1
at $4 h^{-1}$Mpc, and becoming unity at $2h^{-1}$Mpc. This difference
is due primarily to finite volume effects in the GIF simulation, which
is missing some of the large-scale power included in the $512^3$
simulation. This deficit is carried through into the galaxy
correlation functions. It should be noted, however, that this effect
does not invalidate the conclusions of \citeasnoun{meetal} and, if
anything, it slightly improves the agreement between our models and
the APM correlation function on scales greater $\gsim 10h^{-1}$Mpc.

We use the same implementation of the semi-analytic galaxy formation
model described in detail by \citeasnoun{meetal} to populate our
simulations with galaxies. Briefly, for each dark matter halo
identified in the simulation at the redshift of interest, a
Monte-Carlo merger history is generated, based upon the extended Press
\& Schechter formalism \cite{ps74,bcek91,bower91,lc93,coleetal}. A set
of simple, physically motivated rules that describe the galaxy
formation process is then applied to the merger history.  The physics
modelled include:
\begin{itemize} 
\item[(i)] the radiative cooling of virialised gas in halos;
\item[(ii)] the depletion of the resulting reservoir of cold gas by
star formation;
\item[(iii)] the feedback arising from stellar winds and supernovae
that regulate the supply of cold gas;
\item[(iv)] the chemical enrichment of the gas and stars;
\item[(v)] the mergers of galaxies due to dynamical friction.
\end{itemize}
The star formation history of each galaxy is then used to determine
its magnitude in any required band. The effect of dust extinction is
determined using the models of \citeasnoun{ferrara99} as described by
\citeasnoun{coleetal} where full details of the semi-analytic
modelling can be found.

Our fiducial model is essentially the same as that of
\citeasnoun{coleetal}, whose parameters are set by the requirement
that the model should reproduce, as closely as possible, a variety of
properties of the local galaxy population, with particular emphasis
placed on the shape and amplitude of the local, b$_{\rm J}$-band
luminosity function measured in the ESO Slice Project (ESP) by
\citeasnoun{zucca97}. Two cosmological parameters, $\sigma_8$ and
$\Gamma$, differ slightly in this work from the values used by
\citeasnoun{coleetal}. This is simply because Cole et al. had the
freedom to choose these values (they selected $\sigma_8=0.93$ and
$\Gamma=0.19$) whereas we are restricted to using the values employed
in the GIF $\Lambda$CDM simulation.  These small differences have no
significant effect on any of the results presented here.  Since
\citeasnoun{coleetal} used the Press-Schechter mass function for dark
matter halos, whereas we use the mass function produced by the GIF
$\Lambda$CDM simulation, we altered slightly the star formation
efficiency, $\epsilon_\star$, in the model in order to obtain as good
a match to the luminosity function as Cole et al.  did. Specifically,
we set $\epsilon_\star=0.01$, whereas \citeasnoun{coleetal} chose
$\epsilon_\star=0.0067$

The semi-analytic model identifies the most massive galaxy in each
halo as the central galaxy, onto which radiatively cooling gas is
focussed.  This galaxy is placed at the centre of mass of the
simulated halo, and is given the peculiar velocity of the centre of
mass.  Satellite galaxies in the halo are assigned the position and
peculiar velocity of randomly chosen dark matter particles within the
halo.  In this way, satellite galaxies always trace the dark matter
within a given halo.

The technique that we employ differs in two important respects from
that used by \citeasnoun{kauff98a}.  Firstly, Kauffmann et al. extract
the full merger history of each dark matter halo directly from the
N-body simulation. Although this approach has the advantage that the
evolution of any individual galaxy can be followed through the
simulation, the properties of the merger histories generated by the
Monte-Carlo method are statistically equivalent to those obtained
directly from the N-body simulation (but have superior mass
resolution). In particular, the merger histories in the N-body
simulation are found not to correlate with environment, in agreement
with the assumptions of the Monte-Carlo approach
\cite{somerville98,lemson99}.  As a result, any statistical properties
such as the correlation function, are not affected by the differences
between the two approaches.  Secondly, Kauffmann et al. adjusted the
parameters of their semi-analytic model to match the zero-point of the
Tully-Fisher relation between spiral galaxy luminosity and rotation
speed rather than the observed luminosity function as we do. As a
result, some of their models give a poor match to the luminosity
function.  \citeasnoun{meetal} have argued that a good match to the
luminosity function is required for a robust estimate of the
correlation function which, furthermore, for $\Lambda$CDM provides a
good match to the data.

Our semi-analytic model also follows the evolution of galaxy
morphology, as described in \citeasnoun{baugh96} and
\citeasnoun{coleetal}, a feature that we exploit in \S\ref{sec:morph}
to investigate the dependence of clustering strength on morphology.
In the model, virialised gas cools to form a rotationally supported
disk and quiescent star formation turns this gas into stars.  A
galactic bulge can be built up through the accretion of relatively
small satellite galaxies or as the result of a violent merger between
galaxies of comparable mass.  In the latter case, any material in a
disk is rearranged into a bulge and a burst of star formation occurs
if cold gas is present.  The morphology of a galaxy can thus change
with time and is related to its environment, which determines both the
rate at which gas cools and the merger rate between galaxies.

\section{Redshift Space Distortions}
\label{sec:red}

\begin{figure}
\psfig{file=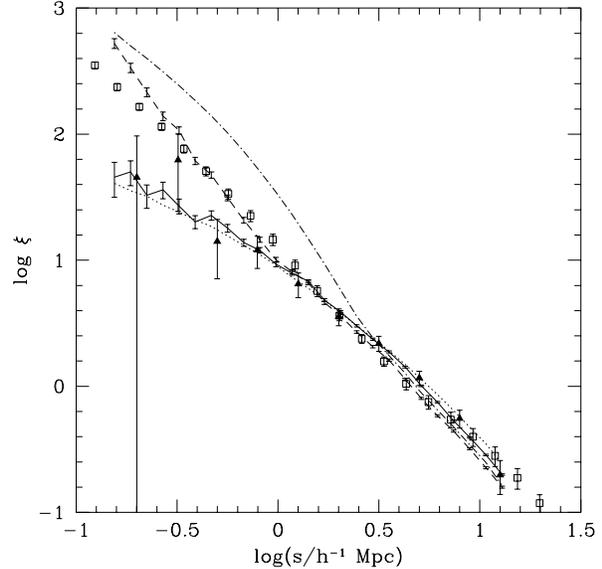,width=80mm}
\caption{Real and redshift-space correlation functions in the
$\Lambda$CDM model.  The redshift-space correlation function of model
galaxies brighter than \mbstartxt\ is shown as a solid line, while the
redshift-space correlation function of dark matter is shown as a
dotted line.  The dashed and dot-dashed lines show the real space
correlation functions for galaxies and dark matter respectively.
Error bars on the galaxy correlation functions are derived from
Poisson statistics. The symbols with error bars are observational
determinations of the galaxy correlation function: triangles show the
redshift-space correlation function of galaxies brighter than
\mbstartxt\ from the ESO Slice Project \protect\cite{guzzo99}, whilst
squares show the real-space galaxy correlation function from the APM
survey \protect\cite{cmbapm}.  }
\label{fig:MBstarred_xi}
\end{figure}

The clustering of galaxies in real space cannot be measured directly
unless a distance indicator is available that does not rely on
redshift.  However, the real space clustering can be inferred
indirectly, by deprojecting either the correlation function measured
in terms of projected galaxy separation or the angular correlation
function \citeaffixed{peebles80}{e.g.}. In the case of noisy projected
measurements, both methods require assumptions about the form of the
two-point correlation function in real space, whilst the second method
also requires knowledge of the redshift distribution of galaxies.

The three dimensional correlation function in redshift-space can be
measured directly, by inferring the radial position of each galaxy
from its redshift, assuming a uniform Hubble flow.  However, galaxies
have peculiar motions in addition to the Hubble flow, induced by
inhomogeneities in the local density field.  This leads to a
modification of the apparent pattern of galaxy clustering, commonly
referred to as ``redshift-space distortions''.  Generally, these
distortions reduce the strength of clustering on small scales, where
the motion of galaxies within virialised structures tends to smear out
galaxy positions inferred from redshifts (creating `Fingers of God'),
and boost the strength of clustering on large scales where galaxies
experience coherent flows into overdense regions and out of underdense
regions \cite{kaiser87,hamilton92}. The redshift-space galaxy
correlation function in our model therefore depends on both the
spatial distribution of the galaxies and their distribution of
peculiar velocities. We will comment on uncertainties that this
introduces in \S\ref{sec:compare}.

The redshift-space clustering of galaxies in the simulation is
straightforward to compute from the position and peculiar velocity of
each galaxy.  Redshift space positions are calculated by adding
$v_{\rm x}/H_0$ to the $x$ coordinate of each galaxy, where $v_{\rm
x}$ is the $x$-component of the peculiar velocity of that galaxy.
This gives the redshift space position (at $z=0$), as viewed by an
observer located at infinity.  From the resulting catalogue we can
compute clustering statistics in the usual way.

In Fig. \ref{fig:MBstarred_xi} we show the two-point correlation
function of galaxies brighter than \mbstartxt\ in redshift space.  The
figure also shows the real-space correlation function for comparison.
The two effects of redshift-space distortion described above are
readily apparent in this figure. In real space the galaxy correlation
function has an almost power-law form between 0.1 and 10$h^{-1}$ Mpc,
with a slope of $\gamma \approx -1.8$.  In redshift-space, the galaxy
correlation function falls below this power-law on small scales,
whereas on scales larger than a few megaparsecs, it rises above it.

\citeasnoun{kaiser87} showed that the amplification in redshift-space
of the two-point correlation function on large scales due to coherent
inflow onto large-scale structures is approximately a factor of $1 +
\frac{2}{3}\beta + \frac{1}{5}\beta ^2$, where $\beta = \Omega
_0^{0.6}/b$ and $b$ is the galaxy bias, assumed to be independent of
scale.  For the model considered here, $\Omega _0 = 0.3$, and the
galaxies are essentially unbiased on large scales.  The amplification
factor is therefore $\sim 1.37$.  Whilst Kaiser's expression assumes
$\Lambda_0=0$, in practice a non-zero cosmological constant makes very
little difference, merely increasing the amplification factor to
$1.39$ \cite{lahav91}.  The value measured for our galaxies at
separations of $10 h^{-1}$ Mpc, where linear theory is still a fair
approximation, is $1.27 \pm 0.04$, in reasonable agreement with the
analytic expectation.

It is remarkable that whilst galaxies and dark matter have very
different two-point correlation functions in real space, on scales
smaller than $3 h^{-1}$ Mpc the redshift-space correlation functions
of galaxies and dark matter are almost indistinguishable. This
implies that galaxies and dark matter have different velocity
dispersions on these scales, and suggests that estimates of galaxy
bias obtained from clustering in redshift space should be treated with
caution.

\subsection{Galaxy peculiar motions}

\begin{figure*}
\begin{tabular}{cc}
Dark matter & Galaxies \\ \psfig{file=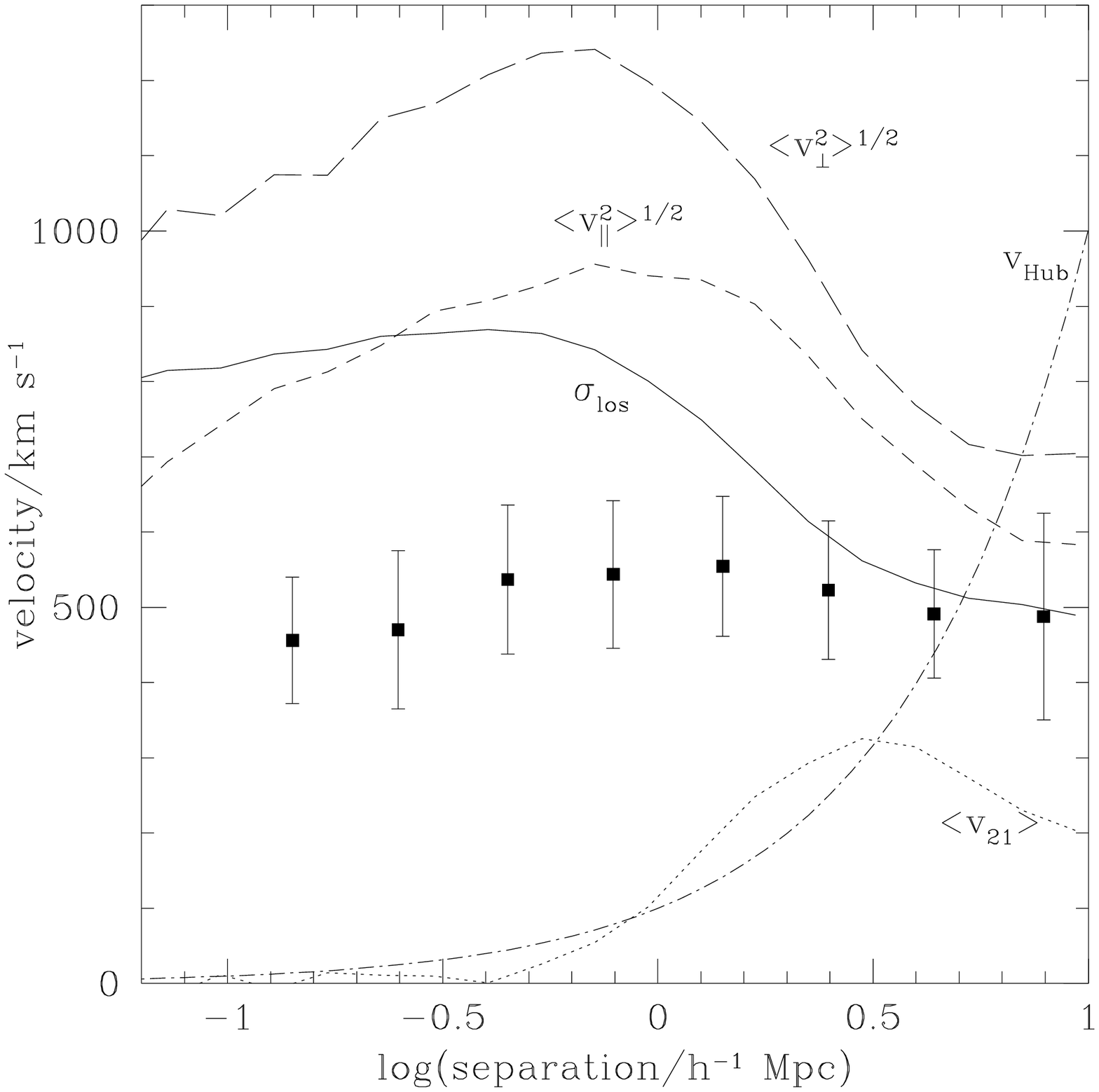,width=80mm} &
\psfig{file=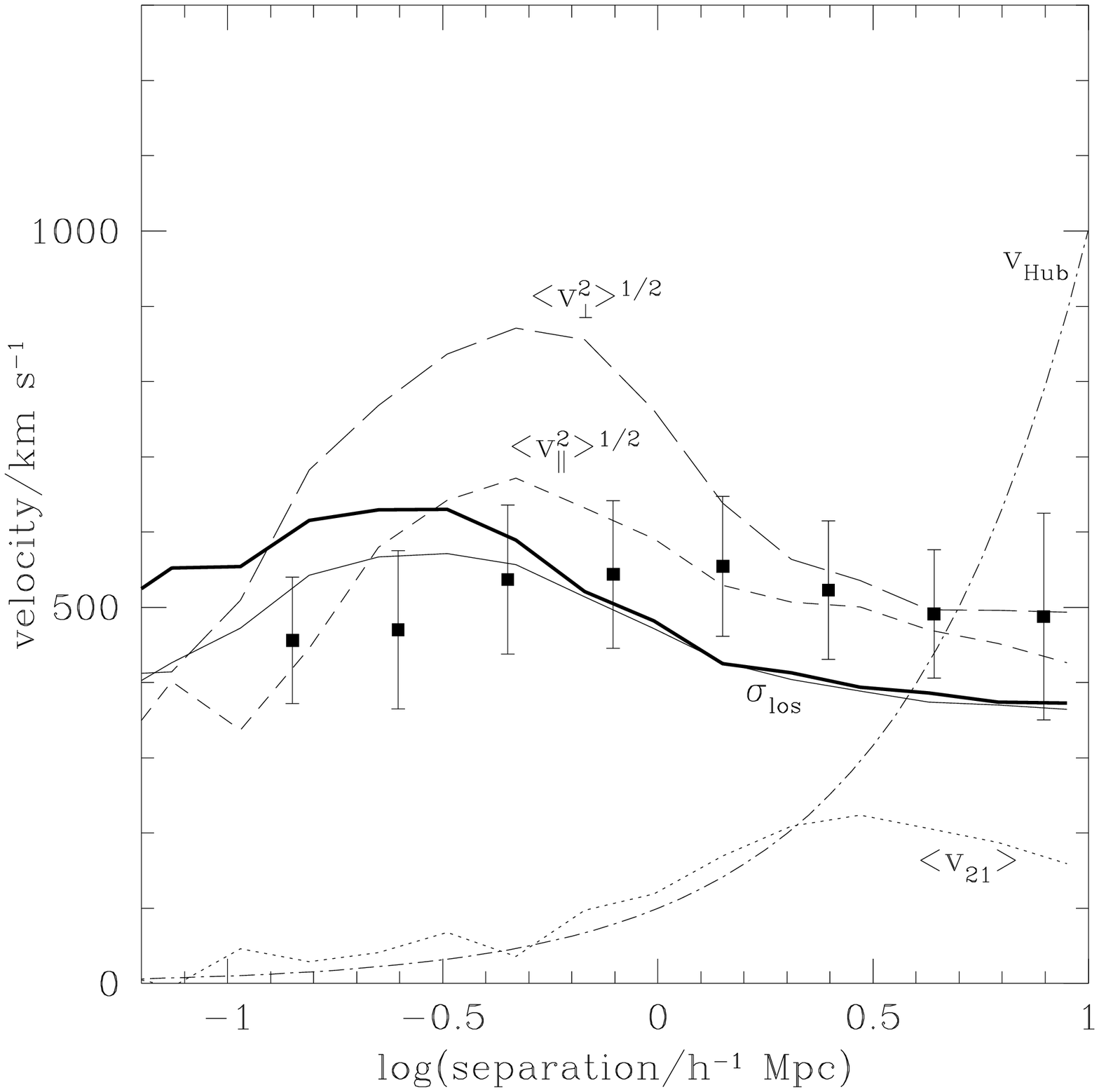,width=80mm}
\end{tabular}
\caption{Pairwise velocities of dark matter particles (left-hand
panel) and galaxies (right-hand panel) measured in the $\Lambda$CDM
simulation. The dotted line is the mean inward radial peculiar
velocity of pairs, $<v_{21}>$; the short-dashed line is the rms
pairwise radial peculiar velocity, $<v_{||}^2>^{1/2}$; the long-dashed
line is the rms pairwise perpendicular peculiar velocity,
$<v_{\bot}^2>^{1/2}$; the solid lines are the rms pairwise
line-of-sight peculiar velocity, $\sigma _{\mathrm{los}}$; and the
dot-dash line is the Hubble expansion given by $v_{\mathrm{Hub}} =
-H_0r$, where $H_0$ is Hubble's constant and $r$ is physical
separation. In the right hand panel the thick solid line shows
$\sigma_{\rm los}$ for a sample with $M_{\rm R}-5\log h\leq -21.5$,
whilst all other lines are for a sample of galaxies with $M_{\rm
B}-5\log h\leq -19.5$. The data points with error bars are taken from
\protect\citeasnoun{jing98} and show the pairwise velocity dispersion,
$\sigma _{12}$, estimated for the Las Campanas redshift survey.  The
error bars are the square root of the sum in quadrature of the errors
derived from the data and $ 1 \sigma$ uncertainties inferred from mock
catalogues by Jing et al.  These points should be compared to the
line-of-sight dispersions, $\sigma_{\rm los}$, for the models (solid
lines). Data points and $\sigma_{\rm los}$ are plotted against
projected separation, $r_{\rm p}$, whilst all other curves are plotted
against true separation, $r$.}
\label{fig:DMpairwise}
\end{figure*}

Pairwise velocity statistics for dark matter and galaxies are plotted
in Fig.~\ref{fig:DMpairwise}. The data points in the figure were
obtained from the Las Campanas Redshift Survey
\citeaffixed{shectman96}{hereafter LCRS,} by \citeasnoun{jing98}.
Following \citeasnoun{arj98}, a quantity that may be compared to these
data is the projected, line-of-sight velocity dispersion given by:
\begin{equation}
\sigma ^2_{\mathrm{los}} (r_{\rm p}) = {\int \xi (r) \sigma ^2_{\mathrm{proj}}
(r) {\mathrm d}l \over \int \xi (r) {\mathrm d}l},
\end{equation}
where $r_{\rm p}$ denotes projected separation, $l$ distance along the
line of sight, $r=\sqrt{r_{\rm p}^2+l^2}$, and the integral is taken
along the line-of-sight between $\pm \infty$, although in practice
convergence is attained with integration limits of $\pm 25 h^{-1}$
Mpc.  The quantity $\sigma ^2_{\mathrm{proj}}$ is the line-of-sight
pairwise dispersion, which is given by:
\begin{equation}
\sigma ^2_{\mathrm{proj}} = {r_{\rm p}^2\left< v^2_{\bot}\right>/2+l^2\left( \left< v_{||}^2\right> -\left< v_{21}\right>^2\right)
\over r_{\rm p}^2+l^2},
\end{equation}
where $\left< v_\perp^2 \right>^{1/2}$ is the rms relative pairwise
velocity perpendicular to the vector connecting each pair, ${\bf
r}_{\rm 12}$, $\left < v_{||}^2 \right>^{1/2}$ is the rms relative
pairwise peculiar velocity along ${\bf r}_{\rm 12}$ and $\left< v_{21}
\right>$ is the mean relative pairwise peculiar velocity along ${\bf
r}_{\rm 12}$, all of which are functions of $r=\left| {\bf r}_{\rm 12}
\right|$.

The line-of-sight velocity dispersion of the {\it dark matter}, shown
in the left-hand panel of Fig.~\ref{fig:DMpairwise}, is significantly
larger than the value for LCRS galaxies on small scales. At $r\sim
0.5h^{-1}$Mpc, the observed value is $535 \pm 100 {\rm kms}^{-1}$,
whereas for the dark matter $\sigma_{\rm los} = 870 {\rm
kms}^{-1}$. For dark matter, we find that the measured $\sigma_{\rm
los}$ from the GIF simulation is in good agreement with that obtained
by \citeasnoun{arj98} from a simulation of a volume approximately 5
times larger than the GIF simulation and from that in the $512^3$
simulation, differing by less than $50 {\rm km s}^{-1}$ on all scales
considered here.

To compare to the LCRS data we have constructed a sample of galaxies
brighter than $M_{\rm R}-5\log h=-21.5$ which corresponds
approximately to the apparent magnitude limit of the LCRS at its mean
depth of $300 h^{-1}$ Mpc. The line-of-sight velocity dispersion of
these {\it galaxies}, shown in the right-hand panel by the thick solid
line, is in reasonable agreement with the data. A sample selected to
have $M_{\rm B}-5\log h\leq-19.5$ has a very similar $\sigma_{\rm
los}$, as shown by the thin solid line in the right-hand panel.

Fig. \ref{fig:DMpairwise} shows that model galaxies display a markedly
different velocity dispersion from the dark matter on all the scales
considered.  This ``velocity bias'' is statistical in origin, and is
unrelated to dynamical friction in clusters, because each galaxy in
our model is assigned the peculiar velocity of a particle belonging to
a dark matter halo in the N-body simulation.  This effect arises
because the galaxy distribution does not constitute a Poisson sampling
of the dark matter distribution.  Firstly, $L_{*}$ galaxies are found
only in dark matter halos more massive than $ \sim
10^{12}h^{-1}M_{\odot}$, whereas a significant fraction of the dark
matter is in smaller mass objects.  Secondly, the number of galaxies
in each halo is not in direct proportion to the halo mass, as would be
required for Poisson sampling \cite{meetal}.  The number of galaxies
per halo determines how well the velocity structure of the halo is
traced by the galaxy distribution. Our results for galaxy velocity
dispersions differ somewhat from those found by \citeasnoun{kauff98a}
using a similar technique. In \S\ref{sec:compare} we explore
thoroughly the causes of these differences.

\begin{figure}
\psfig{file=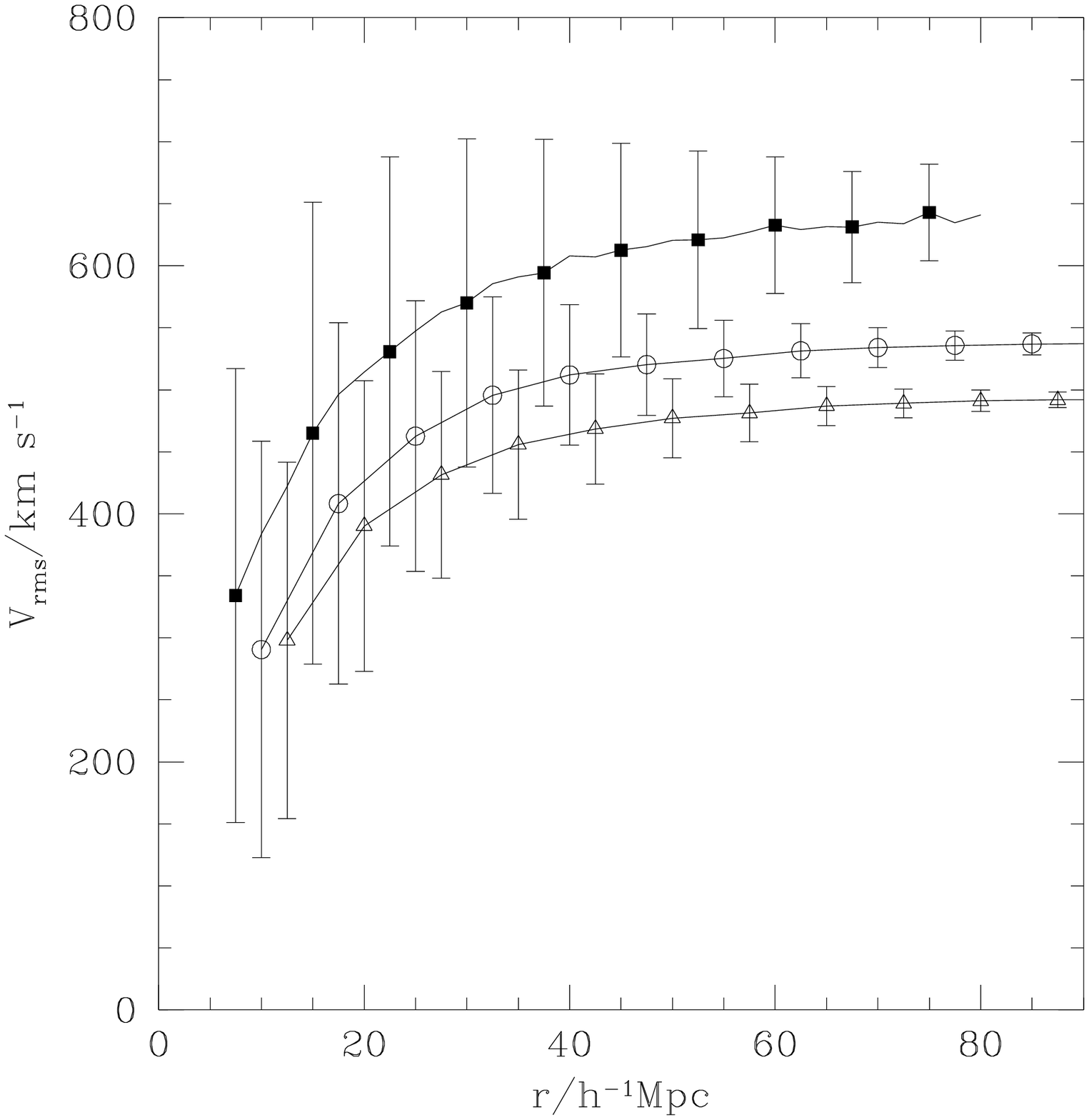,width=80mm}
\caption{The three-dimensional velocity dispersion of individual
galaxies and dark matter in spheres of radius $r$ measured relative to
the mean velocity within the sphere. Filled squares show the velocity
dispersion of the dark matter, whilst open symbols show the velocity
dispersion of galaxies brighter than $M_{\rm B}-5\log h=-19.5$
(triangles) and $M_{\rm B}-5\log h=-18.5$ (circles). The symbols
indicate the mean velocity dispersion, and the error bars show the
{\it rms} scatter in this quantity.}
\label{fig:bulk_flow}
\end{figure}

The kind of biases seen in the pairwise velocity dispersion function
are also present in single galaxy statistics.
Fig. \ref{fig:bulk_flow} shows the three-dimensional velocity
dispersion of galaxies and dark matter within spheres of radius $r$,
measured relative to the mean velocity within the sphere. At large
$r$, this quantity tends to a constant value, the rms peculiar
velocity of individual galaxies. As the figure shows, the dispersion
within spheres is lower for galaxies than for dark matter and,
furthermore, it depends on galaxy luminosity. For bright galaxies, the
velocity dispersion is about 20\% lower than for the dark matter. The
dependence on luminosity arises because the fainter sample contains
relatively more galaxies in the most massive halos which have the
largest velocity dispersions. The figure also shows that the sampling
variance of the velocity dispersion function is very large,
particularly on small scales. For example, for sphere radii of $10
h^{-1}$ Mpc, velocities as low as $\sim 100$ km s$^{-1}$, are within
1-$\sigma$ of the mean dispersion. This reflects the importance of
long wavelength density fluctuations in determining the small-scale
velocity field.

\subsection{Comparison to previous work}
\label{sec:compare}

The pairwise galaxy velocity dispersions that we find are somewhat
lower than those obtained by \citeasnoun{kauff98a} using exactly the
same N-body simulations as us, but a different semi-analytic
prescription. Kauffmann et al found very similar (to within 10\%)
pairwise velocities for dark matter particles and galaxies on all
scales. Thus, unlike ours, their model disagrees with the
\citeasnoun{jing98} data. Since the two semi-analytic treatments
attempt to model the same physics, it is instructive to understand the
reasons for this discrepancy.

\begin{figure}
\psfig{file=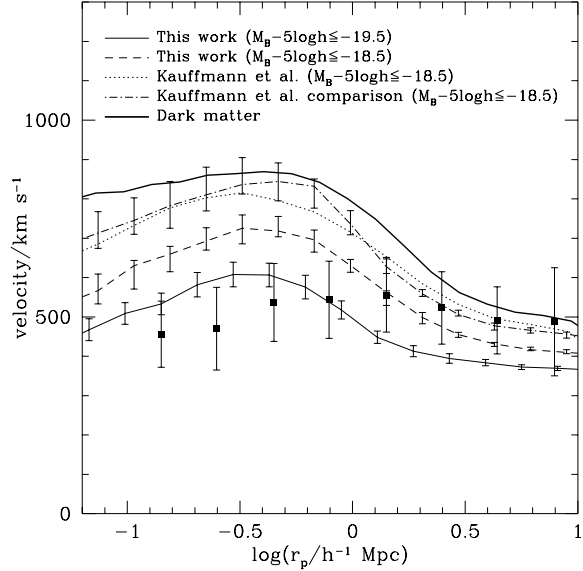,width=80mm}
\caption{Line-of-sight velocity dispersions, $\sigma_{\rm los}$ for:
dark matter (thick solid line); galaxies in our model that are
brighter than $M_{\rm B}-5\log h=-19.5$ (thin solid) and $M_{\rm
B}-5\log h=-18.5$ (dashed line); galaxies in the model of
\protect\citeasnoun{kauff98a} that are brighter than $M_{\rm B}-5\log
h=-18.5$ (dotted line); the Kauffmann et al. galaxies but with
positions and velocities assigned using our placement scheme
(dot-dashed line). The error bars on these lines reflect sampling
uncertainties in the simulated volume as discussed in the
text. Squares with error bars are estimated from the Las Campanas
Redshift Survey \protect\cite{jing98}.}
\label{fig:sigcompare}
\end{figure}

The first, and most obvious difference is that \citeasnoun{kauff98a}
considered galaxies one magnitude fainter than we do and the pairwise
velocities are higher for fainter galaxies (because fainter samples
consist of proportionally more satellite galaxies in clusters than
brighter cuts). In our model, the line-of-sight velocity dispersion,
$\sigma_{\rm los}$, of galaxies brighter than $M_{\rm B}-5 \log h =
-18.5$ is approximately 150~km~s$^{-1}$ higher than that for galaxies
one magnitude brighter. However, as may be seen in
Fig.~\ref{fig:sigcompare}, even for galaxies of the same luminosity,
there is still a discrepancy between our results and those of
Kauffmann et al. There are three effects that could give rise to this
discrepancy: (i) different assignments of peculiar velocities and/or
positions to galaxies in the same halo; (ii) differences in the
frequency with which the two semi-analytic models populate a halo of
given mass with a particular number of galaxies; and (iii) statistical
variations in the formation histories of galaxies in halos of a given
mass. We have carried out several tests of these effects.

\begin{figure}
\psfig{file=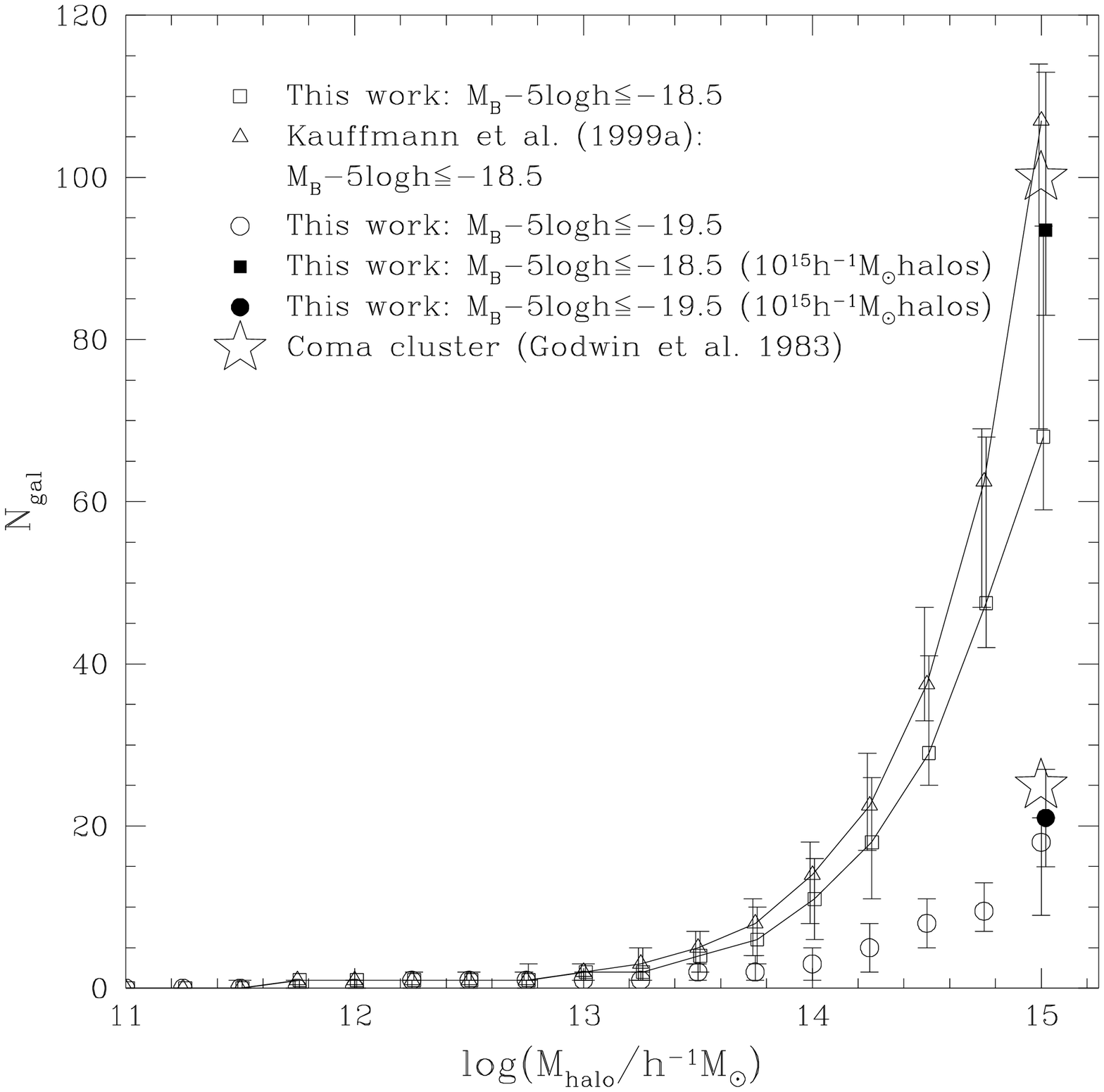,width=80mm}
\caption{The number of galaxies per halo brighter than $M_{\rm
B}-5\log h=-18.5$ as a function of halo mass in our model (squares)
and that of \protect\citeasnoun{kauff98a} (triangles). Also shown is
the number of galaxies per halo brighter than $M_{\rm B}-5\log
h=-19.5$ as a function of halo mass in our model (circles). Filled
symbols show the corresponding predictions from our model for an
independent sample of twenty $10^{15}h^{-1}M_\odot$ halos. The symbols
show the median of the distribution whilst error bars indicate the
10\% and 90\% intervals of the distribution. The number of galaxies
brighter than $M_{\rm B}-5\log h=-18.5$ and $-19.5$ within the virial
radius of the Coma cluster as found by \protect\citeasnoun{godwin83}
are indicated by stars.}
\label{fig:ngalmhalo}
\end{figure}

The results of the tests are shown in Fig. \ref{fig:sigcompare}. The
thick solid line shows $\sigma_{\rm los}$ for the dark matter in the
GIF simulation, whilst the thin solid line shows $\sigma_{\rm los}$
for galaxies brighter than $M_{\rm B}-5\log h\leq -19.5$ in our
model. In order to estimate the variance arising from different galaxy
formation histories, we have generated ten realisations of our model
using different random number sequences when constructing dark matter
merger trees and selecting random halo particles on which to place the
galaxies. The variation in $\sigma_{\rm los}$ between these ten models
is an underestimate of the true statistical uncertainty because the
population of dark matter halos is, of course, identical in all ten
realizations. The dashed line shows $\sigma_{\rm los}$ for galaxies in
our model brighter than $M_{\rm B}-5\log h\leq -18.5$, with $1\sigma$
error bars derived in this way. Although $\sigma_{\rm los}$ for these
galaxies is larger than for our brighter sample, it is still biased
relative to the dark matter and is significantly below the Kauffmann
et al. result. Thus, statistical uncertainties and magnitude
selections, although relatively large, do not, on their own, explain
the difference between the two semi-analytic models.

We now consider differences in the way in which the two models assign
positions and velocities to galaxies within a given halo. In our
model, each galaxy is assigned the position and velocity of a randomly
chosen dark matter halo particle (except the central galaxy which is
assigned the position and velocity of the centre of mass of the
halo). In the model of \citeasnoun{kauff98a}, galaxies initially form
on the most bound particle in their halo. If that halo later merges
with a larger one, the galaxy becomes a satellite in the new halo and
remains attached to the dark matter particle on which it
formed. \citeasnoun{diaferio99} have shown that this placement scheme
leads to a dynamical velocity bias, such that blue galaxies in
clusters have a higher velocity dispersion than the cluster dark
matter because they preferentially occupy particles which are in the
process of falling into the cluster for the first time. Such dynamical
biases cannot (by construction) arise in our model.

To test the effect of the different placement schemes, we have taken
the galaxy catalogue of Kauffmann et al. (limited at $M_{\rm B}-5\log
h=-18.5$ and kindly provided by A. Diaferio) and identified the dark
matter halo of the GIF simulation to which each of these galaxies
belongs. We have then assigned positions and velocities to these
galaxies using our own placement scheme. The resulting $\sigma_{\rm
los}$ is shown in Fig. \ref{fig:sigcompare} as the dot-dashed line
(with error bars calculated from ten realisations of the placement
scheme). This may be compared to the $\sigma_{\rm los}$ for the
original Kauffmann et al. catalogue shown as the dotted line. The two
curves are very similar on small scales and the small differences that
there are on scales larger than $1 h^{-1}$ Mpc are, in fact, due to
our choice of always siting one galaxy at the centre of mass of each
halo. (When we place \emph{all} the Kauffmann et al. galaxies on
randomly chosen halo particles the two curves agree extremely well on
all scales.) This test demonstrates that the differences between our
results and those of Kauffmann et al. are not due to the different
ways in which galaxies are identified with dark matter halo particles
in the two models.

We conclude that the differences between our results and those of
Kauffmann et al. must be due to the frequency with which halos of
different mass are populated with a particular number of galaxies in
the two semi-analytic models. This statistic is shown in
Fig.~\ref{fig:ngalmhalo}. Our model and that of Kauffmann et al. agree
well for halos below $\sim 10^{13}h^{-1}M_\odot$, but for higher
masses Kauffmann et al. assign systematically more galaxies to each
halo (typically around 30\% more) than we do. Thus, the catalogue of
Kauffmann et al. gives greater weight to the most massive, highest
velocity dispersion halos to which pairwise statistics are quite
sensitive. As a result, they predict significantly higher velocity
dispersions than we do. Since the $10^{15}h^{-1}M_\odot$ bin contains
only three halos in the GIF simulation, we also show our model
predictions (filled symbols) from an independent sample of twenty
halos of mass $10^{15}h^{-1}M_\odot$ populated using exactly the same
galaxy formation rules (indicated by filled symbols). For comparison,
we have indicated in the figure the number of galaxies brighter than
$M_{\rm B}-5\log h=-19.5$ found in the Coma cluster and the
corresponding number in our model. The virial mass and radius of the
Coma cluster are $0.8 \pm 0.1 \times 10^{15}h^{-1}M_\odot$ and $1.5
h^{-1}$Mpc ($1.2^\circ$) respectively \cite{geller99}. Within a
circular aperture of this radius centred on NGC4874,
\citeasnoun{godwin83} find 100 galaxies brighter than $M_{\rm B}-5\log
h=-18.5$ and 25 galaxies brighter than $M_{\rm B}-5\log h=-19.5$ (at
the fainter magnitude cut a contamination of approximately 5.5 field
galaxies is expected). Although this sample of one cluster cannot be
used to confirm our model, it is consistent with our predicted
distribution of galaxy numbers but also with that predicted by
Kauffmann et al. To differentiate between these two models would
require a larger observational sample of rich clusters. Our model has
been shown to be in good agreement with the observed total
luminosities of groups and clusters determined by \citeasnoun{moore93}
\citeaffixed{meetal}{Figure 16 of}, indicating that it is populating
both groups and clusters with galaxies in a realistic fashion.

Although velocity statistics are sensitive to the numbers of galaxies
placed in high mass halos, we have checked that galaxy correlation
functions (in both real and redshift-space) are much less
affected. For example, the redshift-space correlation function of
galaxies brighter than $M_{\rm B}-5\log h\leq -18.5$ differs by only
10\% on scales of $10h^{-1}$Mpc and by less than 20\% on scales of
$0.3h^{-1}$Mpc when estimated using the number of galaxies per halo
predicted by our model and by that of \citeasnoun{kauff98a}. This is
comparable to the scatter between models with different parameters
found by \citeasnoun{meetal}.

\begin{figure}
\psfig{file=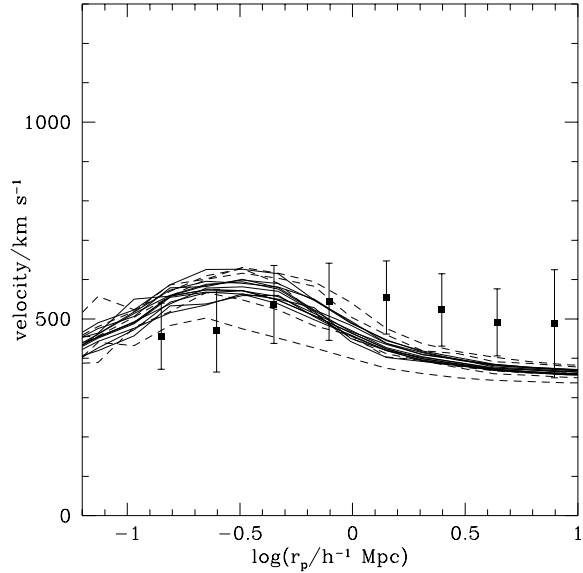,width=80mm}
\caption{The line-of-sight pairwise velocity dispersion, $\sigma_{\rm
los}$, for galaxies with $M_{\rm B}-5\log h\leq-19.5$. The lines show
fifteen variants of our fiducial model in which some of the key
parameters have been varied. The five models shown as dashed lines
would be rejected as being non-realistic, as discussed in
\S\ref{sec:xilum}. The data points with error bars are taken from
\protect\citeasnoun{jing98} and show the pairwise velocity dispersion,
$\sigma _{12}$, estimated for the Las Campanas redshift survey.}
\label{fig:pairwise_rob}
\end{figure}

Fig. \ref{fig:pairwise_rob} shows the effects of varying key model
parameters on the pairwise velocity dispersions. In \S\ref{sec:xilum}
we describe these models further and and explain our criteria for
rejecting models as being non-realistic. Of the fifteen variants
considered, five would be classed as non-realistic, and are shown as
dashed lines. These models show a greater spread in velocity
dispersions than the viable models (solid lines) which have only a
small scatter around the fiducial model. This demonstrates that robust
predictions for velocity dispersions can be made once non-realistic
models are excluded.

\section{The dependence of clustering on galaxy properties}
\label{sec:xitypes}

We now examine the clustering properties of various subsamples
extracted from our original galaxy catalogue, selecting by luminosity,
morphology or colour.

\subsection{Dependence of clustering on luminosity}
\label{sec:xilum}

\begin{figure}
\psfig{file=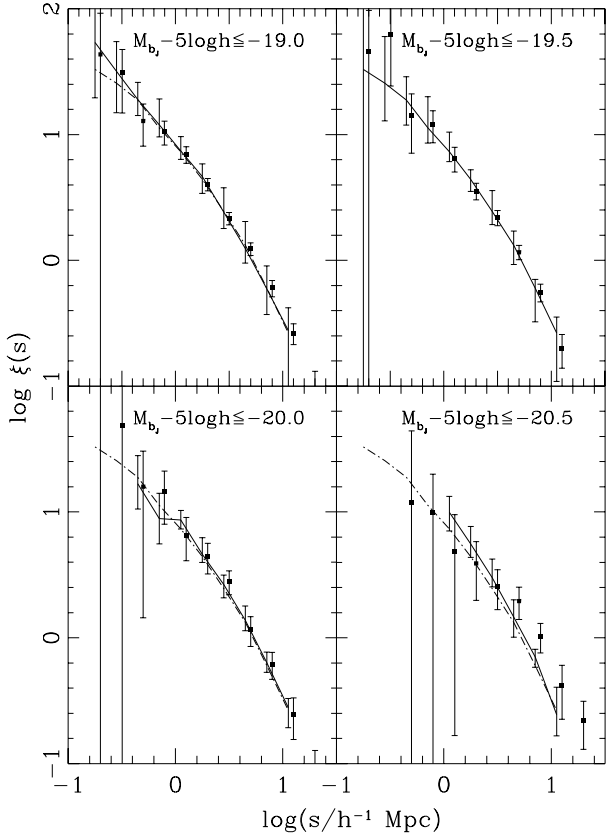,width=90mm}
\caption{Redshift-space correlation functions of galaxies selected by
b$_{\rm J}$-band absolute magnitude (see panels for magnitude
cuts). The lines show the median model correlation function obtained
from fifty randomly chosen volumes equal in size to the corresponding
ESP sample.  The error bars on these lines indicate the 10\% and 90\%
intervals of the distribution of correlation functions.  The model
result for $M_{\rm b_J} - 5 \log h \leq -19.5$ is reproduced, for
reference, in each panel as the dot-dashed line. The filled squares
with error bars show the correlation functions estimated from the
volume-limited samples of the ESP \protect\cite{guzzo99}.  }
\label{fig:ESOmagvar_xi}
\end{figure}

\begin{figure}
\psfig{file=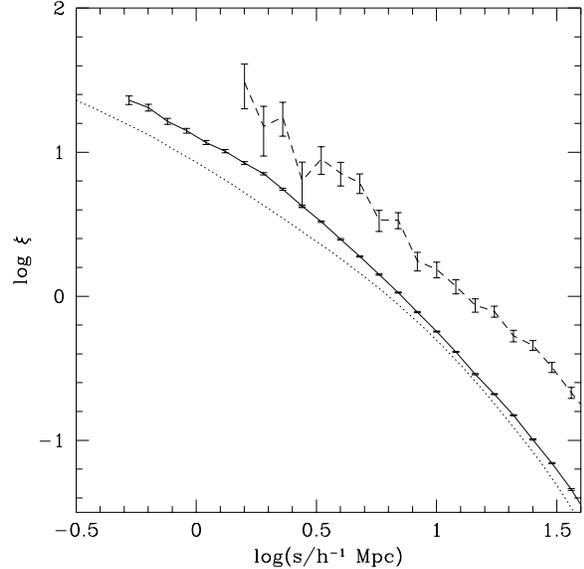,width=80mm}
\caption{Redshift-space correlation functions of galaxies brighter
than $M_{\rm b_J}-5\log h=-20.5$ (solid line) and -21.5 (dashed line)
measured in the $512^3$ simulation. The redshift-space correlation
function of dark matter in this simulation is shown by the dotted
line.}
\label{fig:ESOred_512}
\end{figure}

\begin{figure}
\psfig{file=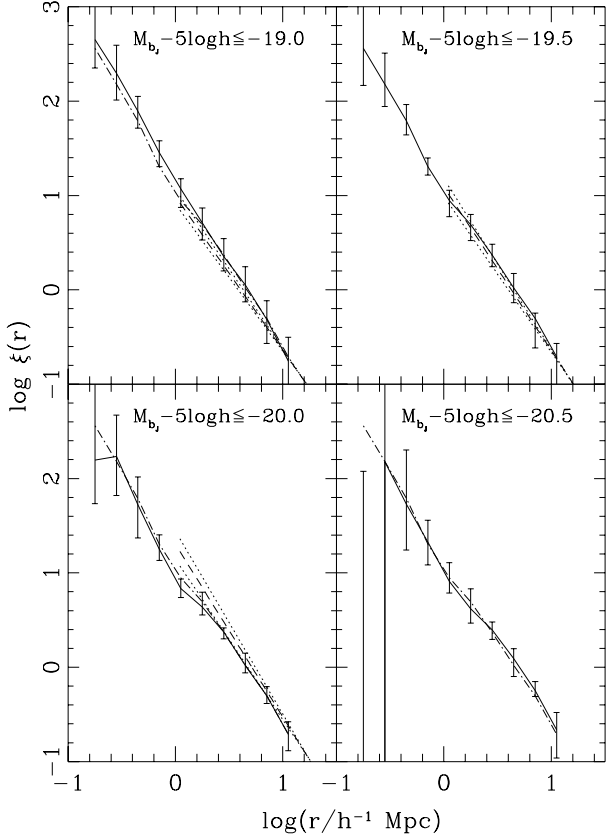,width=90mm}
\caption{ Real-space correlation functions of galaxies selected by
b$_{\rm J}$-band absolute magnitude (see panels for magnitude cuts).
The dashed lines show the power-law fits to the corresponding ESP
survey correlation functions obtained by
\protect\citeasnoun{guzzo99}. Dotted lines track the quoted errors on
these fits. The solid line in each panel shows the median model
correlation function obtained by splitting our sample into fifty
randomly positioned cubes with volume equal to that of the ESP survey
at the same magnitude cut. Error bars on this line indicate the 10\%
and 90\% intervals of the distribution from these same samples.  The
model result for $M_{\rm b_J} - 5 \log h \leq -19.5$ is reproduced,
for reference, in each panel as the dot-dashed line.  }
\label{fig:ESOmagvar_realxi}
\end{figure}

\begin{figure}
\psfig{file=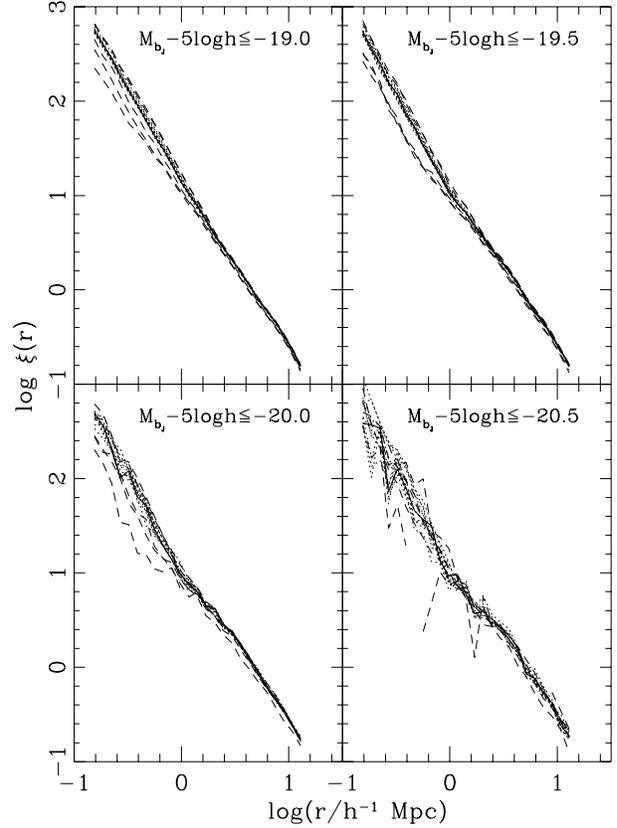,width=90mm}
\caption{Real-space correlation functions of galaxies selected by
b$_{\rm J}$-band absolute magnitude (see panels for magnitude
cuts). The fifteen variant models are plotted either as dotted lines
if they produce a correlation function similar to the standard model
or as dashed lines if they do not.}
\label{fig:ESOmagvar_realxi_rob}
\end{figure}

\begin{figure}
\psfig{file=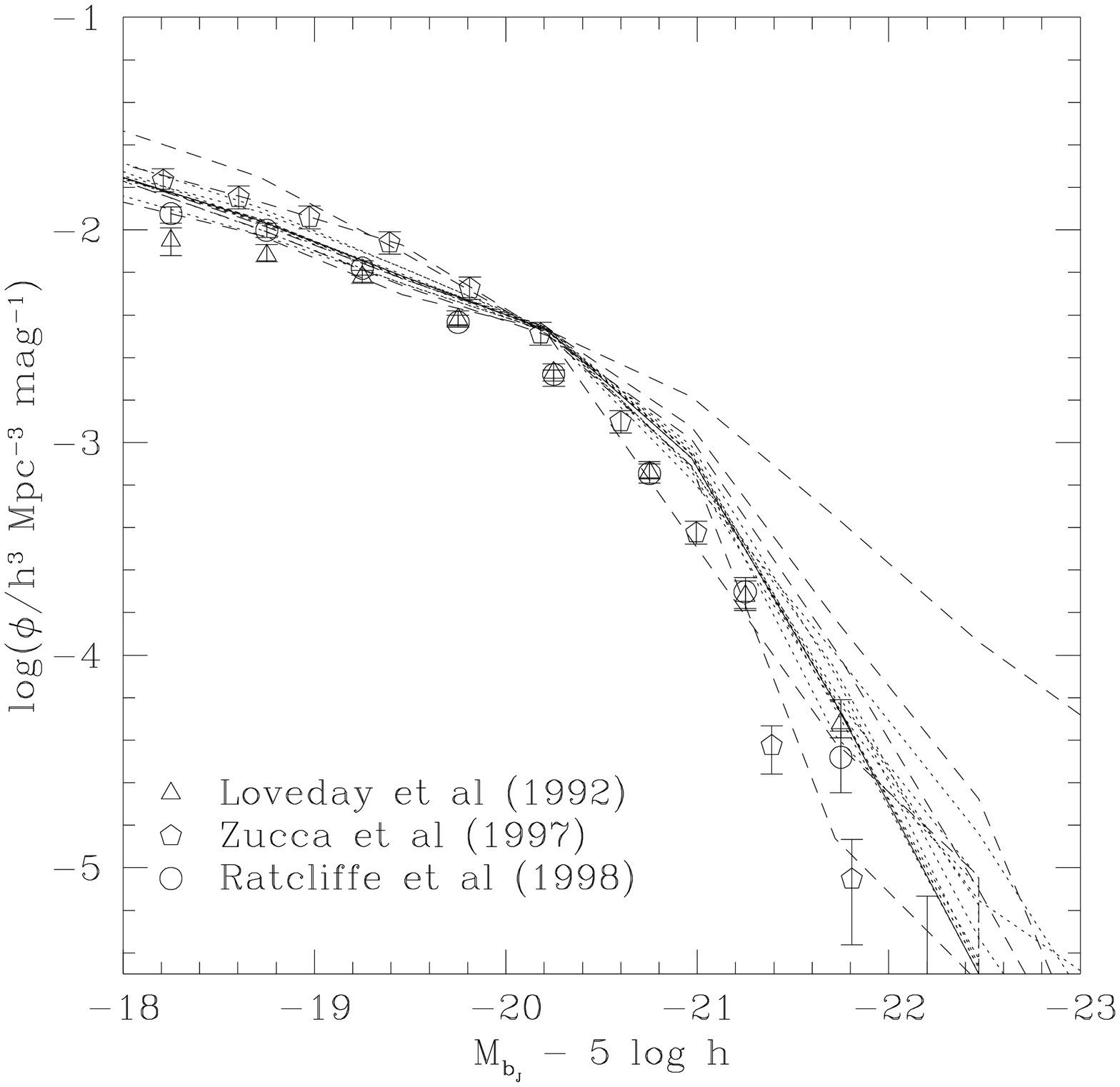,width=80mm}
\caption{The local b$_{\rm J}$-band luminosity function. Symbols with
error bars show various observational estimates of this quantity, as
indicated in the legend. The lines show fifteen variant models along
with our standard model (solid line). The five models which show
significantly different correlation functions are identified by dashed
lines.}
\label{fig:LFB_rob}
\end{figure}

In Fig. \ref{fig:ESOmagvar_xi} we show our predicted redshift-space
correlation functions for four cuts in absolute magnitude, and compare
them with the corresponding correlation functions measured in the ESP
redshift survey by \citeasnoun{guzzo99}. (The space density of
galaxies in our model at each of these magnitude cuts matches closely
that found in the ESP survey, since our b$_{\rm J}$-band luminosity
function is tuned to agree with the ESP survey --- c.f. Fig
\ref{fig:LFB_rob}). The error bars on the model are estimates of the
sample variance, computed in the following manner. For each magnitude
cut, we extract fifty randomly placed cubes from our simulation which
have a volume equal to that of the corresponding volume-limited
samples drawn from the ESP survey. For the ESP survey volume-limited
at $M_{\rm b_J} - 5 \log h \leq -18.5$, $-19.5$ and $-20.5$, our
simulation volume contains 46, 17 and 6 independent survey volumes
respectively. Thus, for the brighter samples our estimates of the
sample variance is likely to be an underestimate of the true value. In
the figure, we plot the median correlation function of these fifty
realisations, and error bars showing the 10\% and 90\% intervals of
the distribution.  As can be seen, the model results are in excellent
agreement with the data over the whole range of scales and magnitudes
shown.

\citeasnoun{guzzo99} argue that for the brightest magnitude cut they
considered, $M_{\rm b_J} - 5 \log h \leq -20.5$, there is evidence for
stronger clustering than for galaxies with $M_{\rm b_J} - 5 \log h
\leq -19.5$.  Unfortunately, this effect is most noticeable on scales
somewhat larger than those we can probe reliably with our simulation.
At $ r \sim 10 h^{-1} $Mpc, the largest scale on which we can reliably
measure the correlation function, we find that galaxies with $M_{\rm
b_J} - 5 \log h \leq -20.5$ have a clustering amplitude that is only
$1.07 \pm 0.10$ times greater than that of galaxies with $M_{\rm b_J}
- 5 \log h \leq -19.5$ (when measured from the full volume of our
simulation). Thus, the results from the GIF simulation are
inconclusive. We have therefore made use of the $512^3$ simulation,
described in \S\ref{sec:desc}, to examine the clustering of very
bright galaxies. The correlation function of galaxies brighter than
$M_{\rm b_J}-5\log h = -20.5$ in this simulation agrees with that of
the same galaxies in the GIF simulation, within the rather large
errorbars (due to the limited number of such bright galaxies in the
GIF volume), except on the largest scales where the different dark
matter correlation functions in the simulations (as discussed in
\S\ref{sec:desc}) introduce a similar difference in the galaxy
correlation functions. As shown in Fig.~\ref{fig:ESOred_512}, we find
that, in the $512^3$ simulation, galaxies with $M_{\rm b_J}-5\log
h\leq -21.5$ have a clustering amplitude on scales above $10 h^{-1}$
Mpc which is approximately 4 times higher than that of galaxies with
$M_{\rm b_J}-5\log h\leq -20.5$. The space densities of $M_{\rm
b_J}-5\log h\leq -20.5$ and -21.5 galaxies are approximately 6 and 200
times lower than that of $L_*$ galaxies respectively. The large
increase in clustering amplitude is due to the fact that these very
bright galaxies are found mostly in halos of mass $\gsim M_*$, for
which the halo bias is a rapidly changing function of halo mass
\cite{mowhite96}.

In Fig. \ref{fig:ESOmagvar_realxi} we show the real-space correlation
function for our models applying the same four absolute magnitude
cuts.  \citeasnoun{guzzo99} estimated the real-space correlation
function by fitting a power-law to the correlation function measured
in terms of projected separation.  We plot their results over the
range of scales used in their fit, $0.5 \le r \le 9.5 h^{-1}$Mpc.
Within the estimated errors from sample variance, our models are in
reasonable agreement with the measurements.  Guzzo et al. claim to
find evidence for a weak luminosity dependence in the real-space
correlation function, with galaxies in the brightest magnitude cut
being the most strongly clustered.  This effect is manifest as an
increase in the correlation length with intrinsic luminosity obtained
from their power-law fits to the correlation function.  The slope of
the ESP correlation function also increases for intrinsically brighter
galaxies. In contrast, we find no significant evidence for a
luminosity dependence of either clustering length or the slope of the
correlation function in our models. The pair-counting errors that
Guzzo et al. estimated for the fitted parameters are largest for the
brightest sample, which contains the fewest galaxies.  Our model
correlation functions, however, indicate that sample variance is
particularly significant for the {\it fainter} catalogues, which
sample smaller volumes.  Taking this sample variance into account
weakens the significance of the effect claimed by Guzzo et al.

To confirm that our model predictions for $\xi(r)$ are robust, we
follow \citeasnoun{meetal} and consider the effect on the correlation
function of galaxies of varying key parameters in our semi-analytic
model, for example, the galaxy merger rate or the baryon
fraction. Each model is constrained to match the ESP b$_{\rm J}$-band
luminosity function at $L_*$ by adjusting the parameter $\Upsilon$,
which gives the ratio of the total mass in stars (including brown
dwarfs) to that in luminous stars. (See \citeasnoun{meetal} for a
description of these variant models.)

In Fig. \ref{fig:ESOmagvar_realxi_rob} we show the real-space
correlation functions obtained from the fifteen models with altered
parameter values. Of the fifteen variants considered, five exhibit
significantly different clustering properties to the fiducial
model. As can be seen in Fig. \ref{fig:LFB_rob}, these models are
amongst the ones that disagree the most with the observed luminosity
function, where these models are shown as dashed lines. This
highlights again the point emphasized by \cite{meetal} that a good fit
to the galaxy luminosity function is a pre-requisite for a robust
determination of clustering statistics in semi-analytic models. The
same five models were shown as dashed lines in
Fig. \ref{fig:pairwise_rob}, were they show a greater variation in
velocity dispersions than models which are a good fit to the
luminosity function. This shows that reproducing the bright end of the
luminosity function is also important in order to make robust
predictions for galaxy velocity statistics.

\subsection{Dependence of clustering on morphology}
\label{sec:morph}

\begin{figure}
\psfig{file=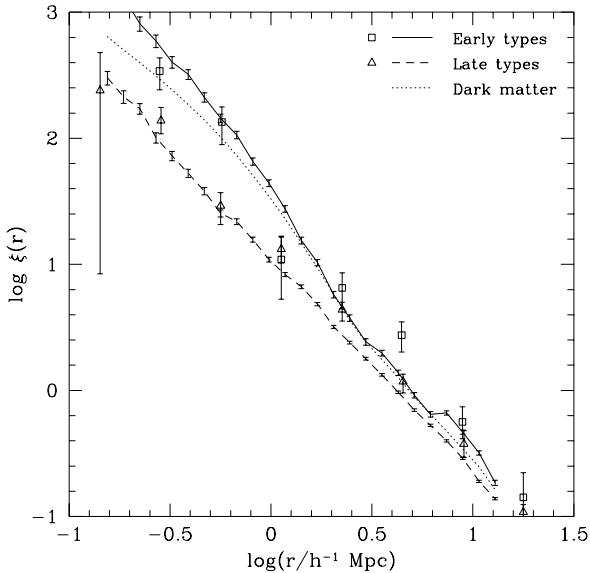,width=80mm}
\caption{Real-space two-point correlation function for galaxy samples
selected by morphological type.  Squares and triangles with error bars
show the observed correlation functions of early and late type
galaxies respectively, with $-20.0 < M_{\mathrm b_J} - 5 \log h <
-19.0$, from \protect\citeasnoun{loveday95}.  Solid and dashed lines
show correlation functions for model galaxies in the same magnitude
range. This sample is subdivided by morphological type as described in
the text. For the model, the lines indicate the correlation function
from the whole simulation volume, which is approximately 1.3 times
larger than the $M_{\mathrm b_J} - 5 \log h < -19.0$ volume-limited
Stromlo-APM survey. The dotted line shows the correlation function of
the dark matter.}
\label{fig:morphxi}
\end{figure}

\begin{figure}
\psfig{file=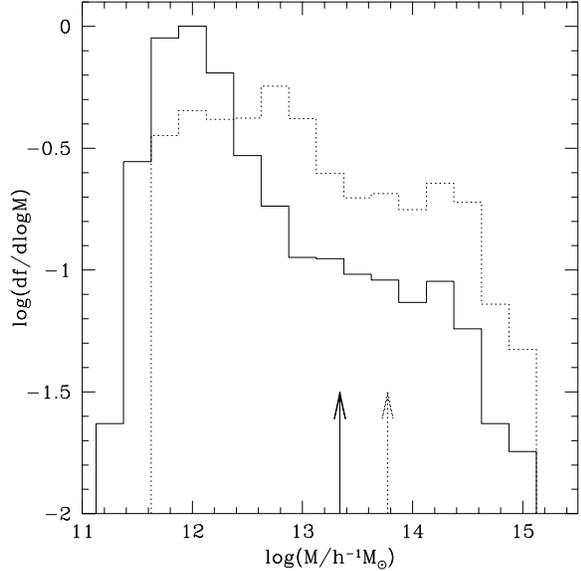,width=80mm}
\caption{The mass function of host dark matter halos, weighted by the
number of galaxies with $-20.0 < M_{\rm b_J} - 5 \log h < -19.0$ per
halo, for late type galaxies (solid line) and early type galaxies (dotted
line) in our model.  The mass functions are normalised by the total
number of galaxies in each sample.  The arrows indicate the mean host
halo mass for each population of galaxies.  }
\label{fig:morphhosts}
\end{figure}

In Fig. \ref{fig:morphxi}, we compare the real-space correlation
functions for galaxies selected by morphological type in our model
with observational data from \citeasnoun{loveday95}.  Although we
restrict attention to galaxies in the same magnitude range used by
Loveday et al., our criteria for morphological classification are
inevitably somewhat different.  Loveday et al. classified galaxies by
eye as either early types (E/SO) or late types (S/Irr).  Our
classification is based on the bulge-to-total luminosity ratio of each
galaxy as measured in dust-extincted B-band light (B/T).  We classify
galaxies with B/T$ < 0.4$ as late type galaxies and those with B/T$ >
0.4$ as early type galaxies.  The observed correlation between
bulge-to-total luminosity ratio and morphological type displays a
large scatter \citeaffixed{baugh96}{see Fig. 1 of }.

Bearing these caveats in mind, the level of agreement between the
observed correlation functions and the model predictions shown in
Fig. \ref{fig:morphxi} is encouraging.  Early type galaxies are the
most strongly clustered on all the scales we consider, as found also
by \citeasnoun{kauff98a}.  The greatest difference occurs on scales
below $1 h^{-1}$ Mpc, where early type galaxies have a clustering
amplitude that is more than 4 times higher than that of late type
galaxies.  On larger scales, the difference in clustering amplitudes
persist, but is less pronounced.

In Fig. \ref{fig:morphhosts}, we plot the host halo mass functions of
late and early type galaxies in our model (weighted by the number of
galaxies per halo). As may be seen, early type galaxies do, on
average, reside preferentially in higher mass environments than late
type galaxies, as speculated by \citeasnoun{loveday95}.  This explains
the larger correlation length of the early types since higher mass
halos are intrinsically more strongly biased \cite{fwde88,mowhite96}
relative to the dark matter distribution as a whole.  It also explains
the greater correlation amplitude of early type galaxies on scales
below $1 h^{-1}$ Mpc, where the correlation function is sensitive to
the number of galaxy pairs inside cluster sized halos \cite{meetal}.
Late type galaxies are rare in these halos and so their pair count on
these scales is small, resulting in a lower correlation amplitude.

\subsection{Dependence of clustering on colour}

\begin{figure}
\psfig{file=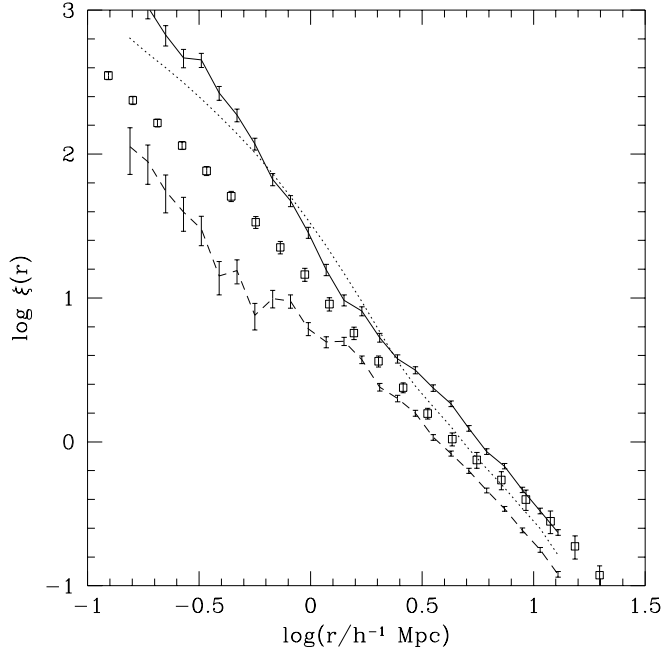,width=90mm}
\caption{The real-space correlation function of galaxies brighter than
$M_{\rm B}-5 \log h = -19.5$ in our model selected by B-R colour. The
solid line is the correlation function of galaxies with B-R$ > 1.3$,
whilst the dashed line is that of galaxies with B-R$ \leq 1.3$. Error
bars on these lines show the Poisson pair counting errors.  The dotted
line shows the correlation function of the dark matter, and the points
with error bars are the real-space correlation function measured from
the APM survey for all galaxies \protect\cite{cmbapm}.}
\label{fig:colxi}
\end{figure}

At present, the only available observational constraints on the colour
dependence of galaxy clustering come from deep, magnitude-limited
samples that cover a significant baseline in redshift.  The technique
used in this paper needs to be extended to include evolution both in
the galaxy population and in the clustering of dark matter to produce
realistic mock catalogues that can be compared with these
observations. We defer a detailed discussion of these techniques and a
thorough comparison to the data to a later paper \cite{benson00}.
However, forthcoming redshift surveys, such as the 2dF
\cite{colless96} and SDSS \cite{gunn95}, will contain large enough
numbers of galaxies with photometry in different bands to allow the
construction of local volume-limited samples subdivided by galaxy
colour.

We present, in Fig. \ref{fig:colxi}, our predictions for colour
dependent clustering of galaxies in real space, measured in the full
volume of our simulation.  Red galaxies are defined to have B-R$ >
1.3$ and blue galaxies B-R$ \leq 1.3$, which matches the cut used by
\citeasnoun{willmer98}. The error bars show the Poisson pair counting
errors in each bin. The clustering of red galaxies is significantly
stronger than that of blue galaxies, which follows from the fact that
the red galaxies are predominantly early types, and so are predicted
to reside preferentially within higher mass halos than the blue
galaxies.

\section{Discussion}
\label{sec:disc}

In this paper, we have combined N-body simulations of the hierarchical
clustering of dark matter with semi-analytic modelling of the physics
of galaxy formation, to probe the relationship between the
distribution of galaxies and the distribution of mass in a
$\Lambda$-dominated cold dark matter universe.

In an earlier paper \cite{meetal}, we studied the correlation function
of galaxies brighter than $L_{*}$ in {\it real-space} (i.e. as a
function of true spatial separation; see also Kauffmann et al
1999a). We found remarkably good agreement between the predictions of
the $\Lambda$CDM model and measurements of the correlation function of
the APM galaxy survey \cite{cmbapm}, over four orders of magnitude in
correlation amplitude.  On small scales, we found the model galaxies
to be {\it less strongly} clustered than (or biased low relative to)
the dark matter. We have now seen that the net effect of small-scale
peculiar velocities is to cancel out this bias: in redshift space the
galaxy and dark matter correlation functions are very similar to one
another. This cancellation arises because the pairwise velocity
dispersions of galaxies and dark matter are different. Fortuitously,
the difference is just sufficient to compensate for the differences in
real-space clustering. Thus, although genuinely biased, the
distribution of galaxies measured in redshift space appears, to a good
approximation, unbiased on small scales.  The pairwise velocity
dispersion of the model galaxies is $\sim(200-300)$ km/s lower than
that of the dark matter over almost two decades in spatial separation
and is in good agreement with recent observational determinations
\cite{jing98}. However, comparison of our results with those of
\citeasnoun{kauff98a} demonstrates the sensitivity of this statistic
to the number of galaxies that populate rich clusters. Both models are
consistent with the available data for the Coma cluster, but produce
line-of-sight velocity dispersions which differ by around 100 km/s.

The physical origin of the offset between the galaxy and dark matter
velocity distributions lies in the way in which galaxies sample the
velocity field of the dark matter. The mass-to-light ratio of halos in
our model is a strong, non-monotonic function of halo mass. Galaxy
formation is most efficient in halos of mass $\sim 10^{12} h^{-1}
M_{\odot}$, and the mass-to-light ratio increases at higher and lower
masses due to long cooling times for the gas and stronger feedback
respectively \citeaffixed{meetal}{see Fig. 8 of}.  As a result, the
number of galaxies per halo does not increase as rapidly as the halo
mass. Thus, when computing the pairwise velocity dispersion, high
velocity dispersion halos are undersampled by galaxies relative to the
contribution of these halos to the velocity dispersion of the dark
matter itself.

We have explored the sensitivity of our theoretical predictions for
peculiar velocities to variations in the galaxy formation
parameters. In \citeasnoun{meetal} we found that predictions for the
galaxy two-point correlation function in a particular cosmological
model are robust to changes in model parameters, provided that the
galaxy luminosity function remains approximately the same. Here, we
find a similar result for the peculiar velocities: models with similar
luminosity functions produce similar results. The amplitude of the
velocity bias depends not only on the luminosity function, but also on
the shape of the power spectrum of density fluctuations and on the
statistics of the occupation of halos by galaxies.  As we showed in
\citeasnoun{meetal}, our (cluster-normalized) $\Lambda$CDM model gives
a good match to the observed luminosity function of galactic
systems. To a large extent, this is the reason why the model also
gives a good match to the observed pairwise velocity dispersion
function (although our neglect of dynamical biases in the galaxy
distribution may affect our results slightly). The differences between
our predicted peculiar velocities and those obtained by
\citeasnoun{kauff98a} from the same N-body simulations, but using a
different semi-analytic model, are simply due to differences in the
way in which the two models populate high mass halos. These, in turn,
reflect differences in the luminosity functions of galaxies and
galactic systems in the two models.

The dependence of clustering on intrinsic galaxy properties provides,
in principle, an interesting test of models of galaxy formation. The
most obvious property to consider is galaxy luminosity. Unfortunately,
the dependence of clustering on luminosity is difficult to measure
from magnitude limited redshift surveys and so the observational
situation is inconclusive.  A weak effect has been claimed, for
example, by \citeasnoun{guzzo99} for galaxies approximately one
magnitude brighter than $L_{*}$. For the bulk of the galaxy
population, we find no evidence for a dependence of clustering on the
intrinsic luminosity of our model galaxies.  This is not surprising in
view of the fact that galaxies of a given luminosity reside in dark
matter halos spanning an appreciable range of masses. We do, however,
predict a strong effect for galaxies that are significantly brighter
than $L_{*}$. Unfortunately, the space density of these galaxies is
too low for their clustering to be measured in existing redshift
surveys. On the other hand, both models and observations agree that
early type or red galaxies are more strongly clustered than late type
or blue galaxies, particularly on small scales. This is a reflection
of the observed morphology-density relation which arises naturally in
hierarchical clustering models \cite{fwed85,kauff96a,baugh96}.

It is worth emphasizing that the galaxy formation model that we have
used in this paper is essentially the same as that discussed at length
in \citeasnoun{coleetal} (except for the small differences mentioned
in \S\ref{sec:desc}) and adopted by \citeasnoun{meetal}.  Cole et
al. fixed parameter values by requiring their model to reproduce a
variety of properties of the local galaxy population, with emphasis
placed on achieving a good match to the local b$_{\rm J}$-band galaxy
luminosity function. This same $\Lambda$CDM model also reproduces the
present day clustering of galaxies in real and redshift space, is in
reasonable agreement with the observed line-of-sight pairwise velocity
dispersion of galaxies, and matches the clustering measured at
intermediate and high redshifts \cite{baugh99}.

An important conclusion of this work is that the statistical
properties of the galaxy distribution can be quite different from
those of the underlying dark matter. A physical approach to modelling
the formation and evolution of galaxies, using the kind of techniques
discussed in this paper and also in \citeasnoun{kauff98a} and
\citeasnoun{meetal}, is therefore imperative if we are to understand
the biases in the way that different galaxies trace the dark
matter. This, in turn, is a pre-requisite for making sense of the
unprecedented amount of information about the galaxy distribution in
the low and high redshift Universe that will shortly become available
from forthcoming redshift surveys.

\section*{Acknowledgements}

AJB, SMC, CSF and CGL acknowledge receipt of a PPARC Studentship,
Advanced Fellowship, Senior Fellowship and Visiting Fellowship
respectively. CSF further acknowledges receipt of a Leverhulme
Fellowship. This work was supported in part by a PPARC rolling grant,
by a computer equipment grant from Durham University and by the
European Community's TMR Network for Galaxy Formation and
Evolution. We acknowledge the Virgo Consortium and the GIF
collaboration for making available the GIF simulations for this study
and the VIRGO Consortium for providing the other N-body simulations
used in this work. We thank Simon White and Nigel Metcalfe for many
valuable comments, and the referee Antonaldo Diaferio both for
supplying his galaxy catalogue and for providing critical and useful
comments which improved the clarity of this paper.

\end{document}